\documentclass[10pt,twocolumn,notitlepage]{article}

\usepackage{epigraph} 
\usepackage[affil-it]{authblk}
\usepackage{dcolumn} 
\usepackage[stable]{footmisc} 
\usepackage{blkarray}
\usepackage[font=small]{caption}
\usepackage{amsmath}
\usepackage[margin=1.7cm]{geometry}
\usepackage{tcolorbox}
\usepackage{hyperref}
\usepackage{float}
\usepackage{graphicx}
\usepackage{subcaption}

\usepackage{amssymb}
\usepackage{enumitem}

\usepackage{tikz, pgfplots, tkz-base, tkz-fct}
\pgfplotsset{compat=newest}

\usepackage{xcolor}
\hypersetup{
    colorlinks,
    linkcolor={black},
    citecolor={black},
    urlcolor={blue!30!black}
}

\title{Accessibility: A Generalization of the Node Degree
\\ (A Tutorial)}

\author{Alexandre Benatti$^1$}
\author{Luciano da F. Costa$^1$$^*$}
\affil{$^1$S\~ao Carlos Institute of Physics, University of S\~ao Paulo, S\~ao Carlos, SP, Brazil\\
$^*$ luciano@ifsc.usp.br}

\begin{document}

\twocolumn[ \begin{@twocolumnfalse} \maketitle \begin{abstract}
Robust and comprehensive characterization of the topological properties of complex networks requires the adoption of several respective measurements, among which the node degree has special importance. In the present work, we provide an introduction to one of these measurements, namely the accessibility of a node, which can be understood as a generalization of the concept of node degree not only to incorporate successive neighborhoods of that node, but also to reflect specific types of dynamics unfolding in the network. After discussing the node degree and its hierarchical extension, we present the concepts of random walk, entropy, and then the accessibility. Several examples of its numeric calculation are provided, as well as some experimental results indicating that it can effectively complement the information provided by other topological measurements of four types of complex networks, namely Erdős–Rényi, Watts-Strogatz, Barabasi-Albert, and Geometric. We also describe how a recently developed toolbox can be used for the calculation of accessibility in relatively large networks.
\end{abstract} \end{@twocolumnfalse} \bigskip
]
\maketitle

\setlength{\epigraphwidth}{.49\textwidth} \epigraph{``D'una citt\`a non godi le sette o le settantasette meraviglie, ma la risposta che d\`a una tua domanda.''}{Le Citt\`a Invisibili (Italo Calvino)}

\section{Introduction}

Complex networks are characterize by the intricacies (or lack of it) of the properties of their topologies, as reflected in several respective measurements~\cite{da2018complex}. In particular, the adjective
\emph{complex} in complex networks has typically indicated how much the topology of the network in question departs from a regular graph, characterized by all nodes having exactly the same degree or, when stochasticity is to be considered, a uniformly random network, whose nodes present similar degrees.  

Though the degree is possibly the most important topological measure, the specification of the degree of each node in a given network is by no means enough to completely represent or specify that network.  Indeed, several networks can be constructed with $N$ nodes having exactly the same degree but differing respectively to other topological properties (e.g.~\cite{da2018complex}). One reason accounting for this phenomenon is that the degree of a node in a network is only defined by the local topology of the network around that node.  More specifically, the degree of a node corresponds to the number of neighbors (or connections) it has.  From another perspective, we can also say that the representation of a network in terms of the degree of all its constituent nodes provides a ``degenerate'' mapping of that network, in the sense of being non-invertible, therefore implying loss of information about the original structure.

\begin{figure}[!htpb]
  \centering
     \includegraphics[width=0.45\textwidth]{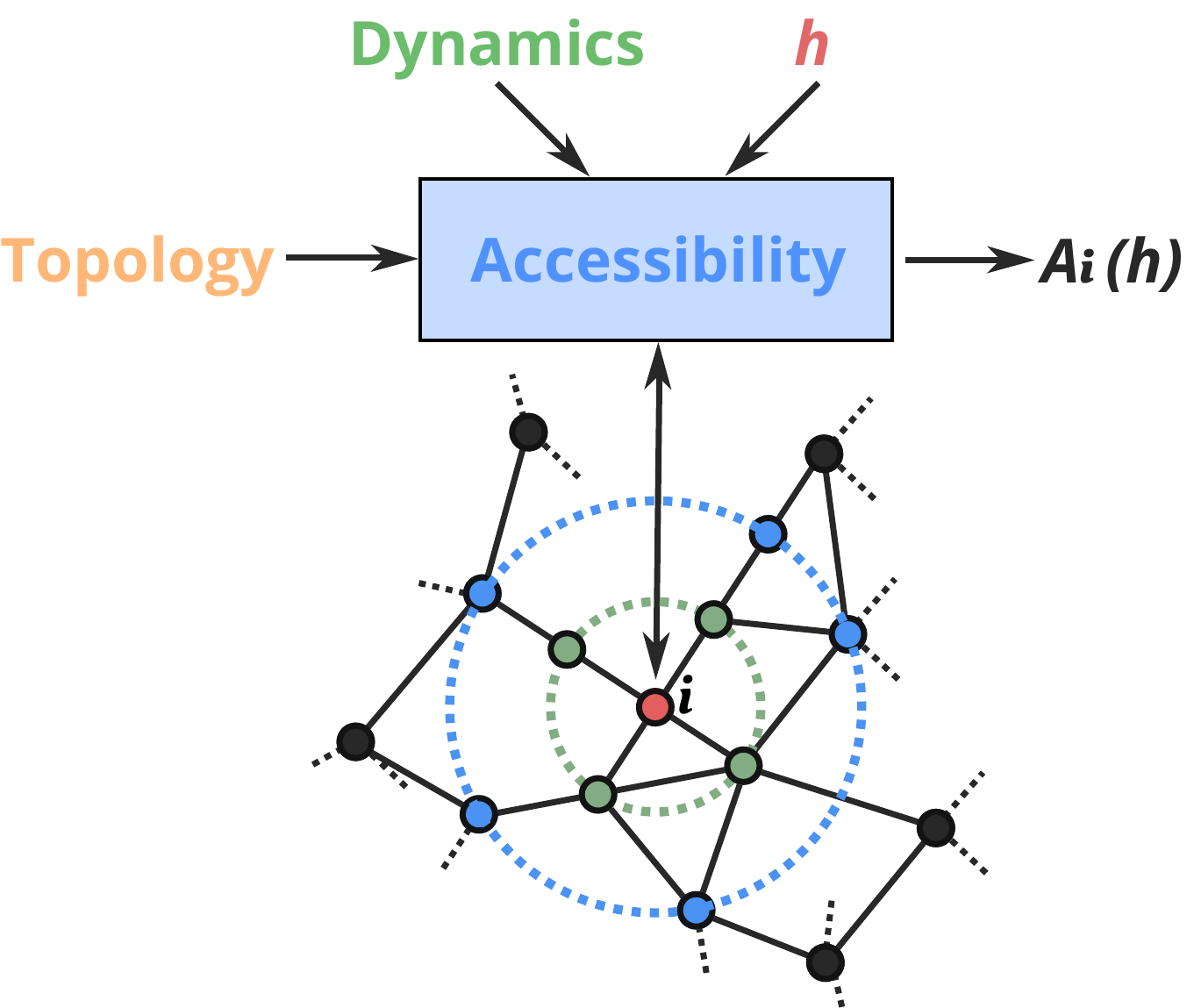}
   \caption{The main aspects underlying the definition of the accessibility of a reference node degree (shown in red) include the adoption of a dynamics of interest, the definition of one or more topological scales (order $h$), and the consideration of the network topology along successive neighborhoods of the reference node (the first two of them being represented in green and blue).}
  \label{fig:scheme}
\end{figure}

The adoption of additional topological measurements of a network contributes to making a mapping less degenerated. For instance, in case the adjacency matrix of a network is taken as its respective set of measurements, it will provide an invertible mapping.  However, this type of measurement depends on the labeling of nodes, therefore implying in solving the isomorphism problem, which can computationally prohibitive. 

There seems to be no definitive result regarding the identification of the smallest set of topological measurements, not dependent of isomorphism, allowing an invertible representation of a network through a respective mapping into a measurement space.  One of the involved complications is that there is a virtually \emph{infinite} set of possible topological measurements of complex networks, some of which have been revised in~\cite{costa2007characterization}.

Some network measurements other than the node degree have more frequently adopted (e.g.~\cite{costa2007characterization}), including the clustering coefficient, shortest path between nodes, assortativity, betweenness centrality, and matching index, among many others. One family of complementary measurements of the topology of a given network that is of particular interest for the present work are those features more closely related to the node degree. These include, but are not limited to, the hierarchical degree, accessibility, and symmetry.

Briefly speaking, the hierarchical degree allows for expanding the topological scale of the degree in order to take into account subsequent neighborhoods. The accessibility expands further the concept of the hierarchical degree to incorporate dynamic properties associated with networks, allowing probabilities to be defined between each node and nodes belonging to several neighborhoods. In particular, the accessibility measurement has been found to be closely related to the borders of a given network, which would correspond to the set of nodes presenting the smallest accessibility values (e.g.~\cite{travenccolo2009border}).

The present work aims at providing a hopefully accessible introduction to the accessibility measurement of a node in a complex network. As a preparation, the concepts of traditional node degree, hierarchical degree, random walks are revised and illustrated first.  Then, the accessibility measurement is presented, illustrated, and compared to some other measurements in terms of Pearson correlation coefficients.  The possibility to use the accessibility to define and obtain the borders of networks is then discussed, followed by a mini-review of some of the many applications of the accessibility to be found in the literature.

\section{Node Degree: A Specially Important Measurement}

The simplest manner to define the node degree probably is as ``the number of edges connected to a node''. Though, we also provide in the following a more formal mathematical definition of this measurement in terms of the adjacency matrix $A$ representing the network of interest. 

The adjacency matrix has dimension $N \times N$, where $N$ is the number of network nodes. The element $a_{i,j}$ of $A$ corresponds to the number of connections between $i$ and $j$, with $a_{i,j}=0$ meaning that these nodes are not connected. On non-directed networks, $A$ will be a symmetric matrix, whereas for directed networks, it is possible that $a_{i,j} \neq a_{j,i}$.

The degree for a node indexed by $i$ in a non-directed network is:
\begin{equation}
    k_i = \sum_j{a_{i,j}},
\end{equation}
where $a_{i,j}$ stands for a generic element of the matrix $A$. 

\begin{figure}[!htpb]
  \centering
     \includegraphics[width=0.45\textwidth]{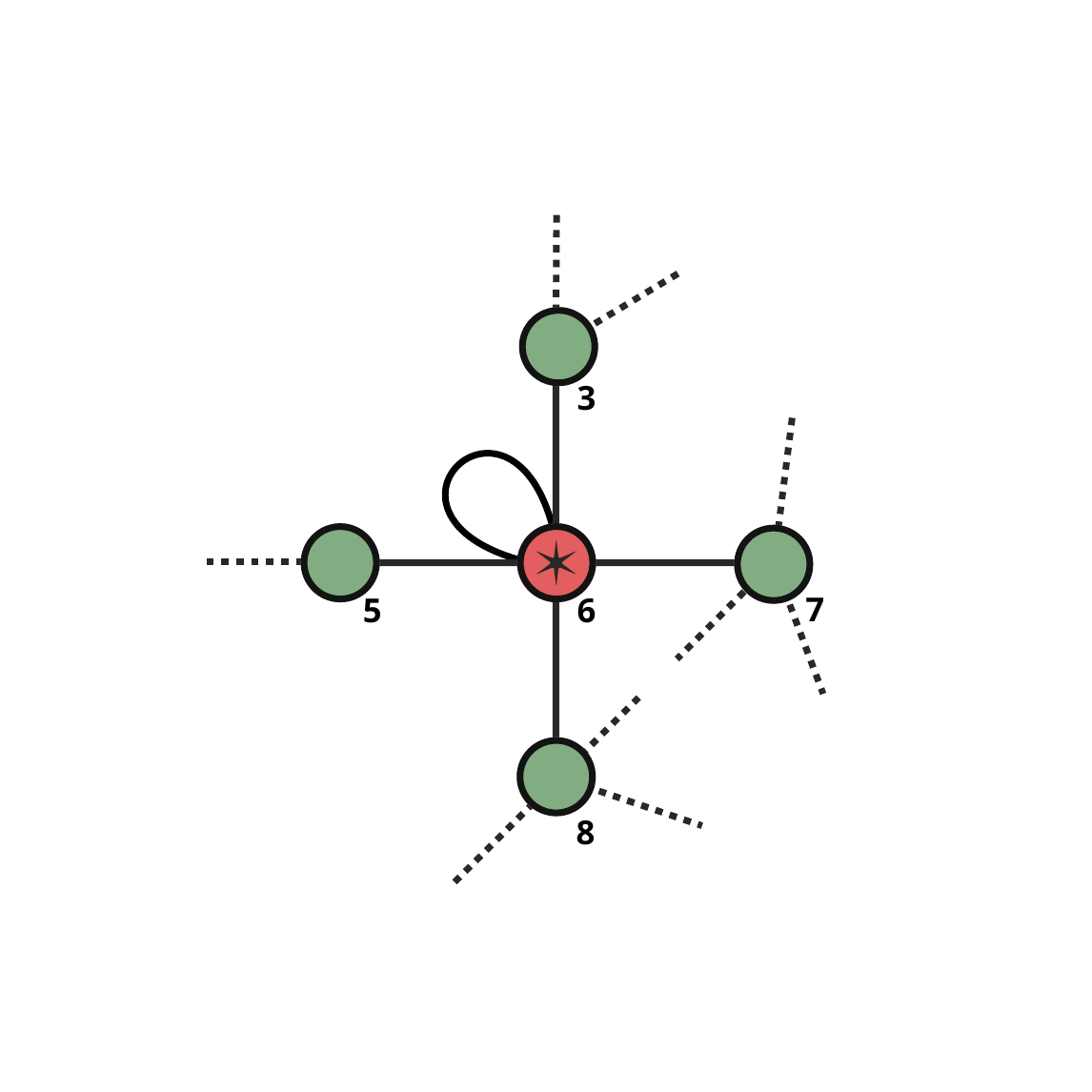}
   \caption{The degree of a given network node. The degree of node $6$ (highlighted) is equal to $k_6=5$ because it makes five connections.}
  \label{fig:Ex_degree}
\end{figure}

Figure~\ref{fig:Ex_degree} shows an example illustrating the concept of degree of a node belonging to a simple network.

In directed networks, each node has both its out- and in-degrees. The former corresponds to the number of edges outcoming from a node ($k_i^{out}$). The in-degree is the number of edges that enter a node ($k_i^{in}$). The total degree of the node therefore corresponds to the sum of in-degree and out-degree:
\begin{equation}
    k_i = k_i^{out} + k_i^{in}.
\end{equation}

Given that each network node has its own degree (or in- and out-degree
in the case of directed networks), frequently the histogram of the degree is used to characterize the topology of a network.   

\begin{figure}[!htpb]
  \centering
     \includegraphics[width=0.45\textwidth]{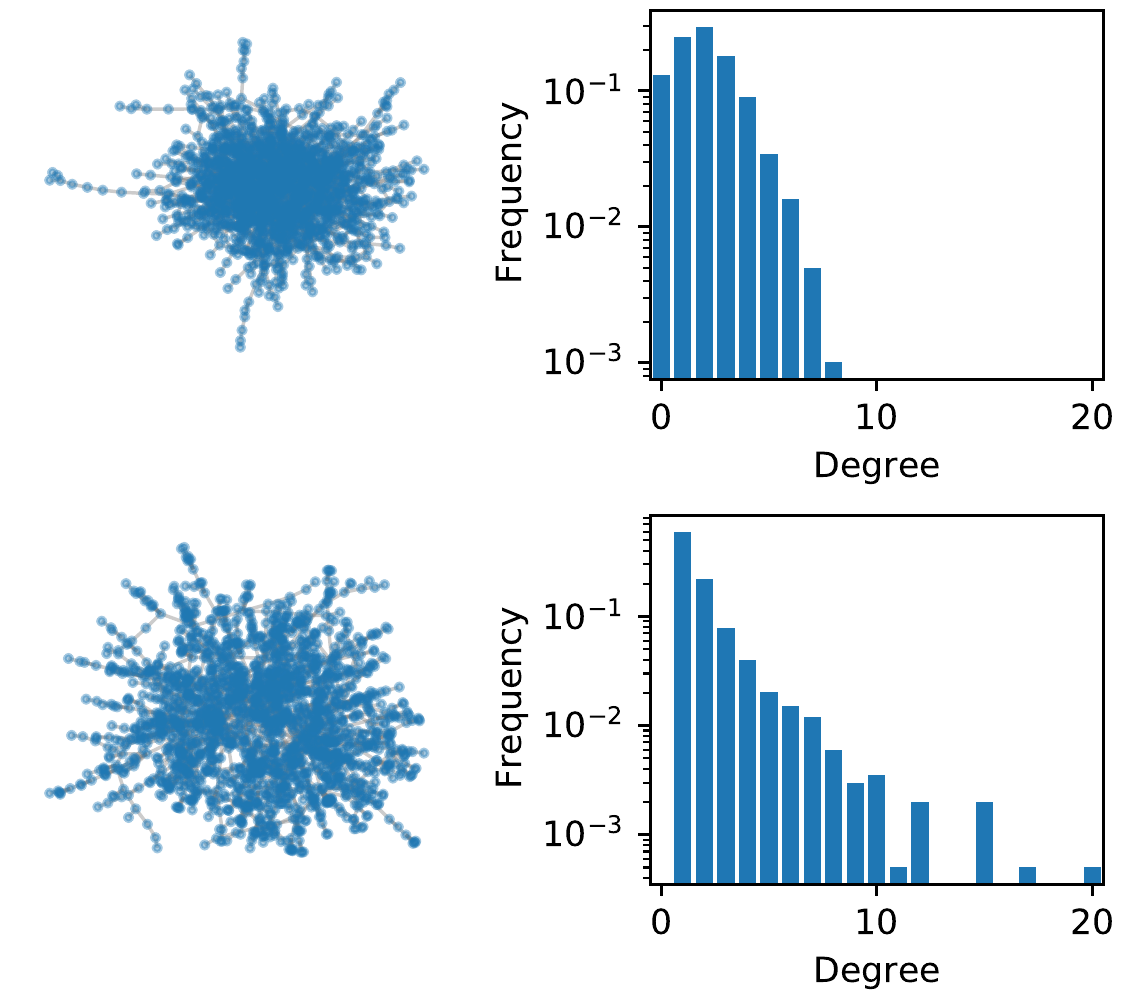}
   \caption{Example of two networks topology (on left) and the respective degree histogram distribution (on right). Both networks have $2,000$ nodes and an average degree equal to $2.0$.}
  \label{fig:deg_hist}
\end{figure}

The node degree is particularly important because it provides a good indication of the local connectivity of a network, as illustrated in Figure~\ref{fig:deg_hist}. In addition, the node degree has been found to be often correlated to several other measurements of the network, as well as being also related to important dynamical properties~\cite{valente2008correlated,valente1998integration}.

However, despite the particular importance of the node degree, the its specification for each of the nodes of a given network is, in general, not enough to provide a complete respective representation, in the sense of being impossible to recover the original network only from this information (e.g.~\cite{da2018complex}). For this reason, several other  measurements need to be be typically incorporated in order to characterize the topology of complex networks in a more complete manner (e.g.~\cite{costa2007characterization}).

Among the several existing topological measurements, the hierarchical degree, accessibility, and symmetry are directly related to the concept of node degree.  In fact, they correspond to successive generalizations of that original concept, as will be discussed in the remainder of the following sections.

\section{Hierarchical Degree}

One manner to generalize the node degree is by considering not only the first neighbors, but also nodes belonging to successive neighborhoods (also called layers). The hierarchical degree of a node $i$ can be defined as the number of links between the nodes that are in the layer $R_{h-1}^i$ and $R_{d}^i$ with node $i$ as reference~\cite{da2006hierarchical}. The layer $R_{h}^i$ is defined as containing all nodes in network that are at minimal distance of $h$ edges, starting from $i$. We refer to $h$ as the \emph{order} of the hierarchical degree.

The hierarchical degree of node $i$ , considering the order $h$, can be expressed by the following equation:
\begin{equation}
    K_i(h) = \sum_{j \in R_{h}^i} \, \sum_{l \in R_{h-1}^i} a_{l,j},
\end{equation}
where $a_{lj}$ are elements of adjacency matrix.

\begin{figure}[!htpb]
  \centering
     \includegraphics[width=0.38\textwidth]{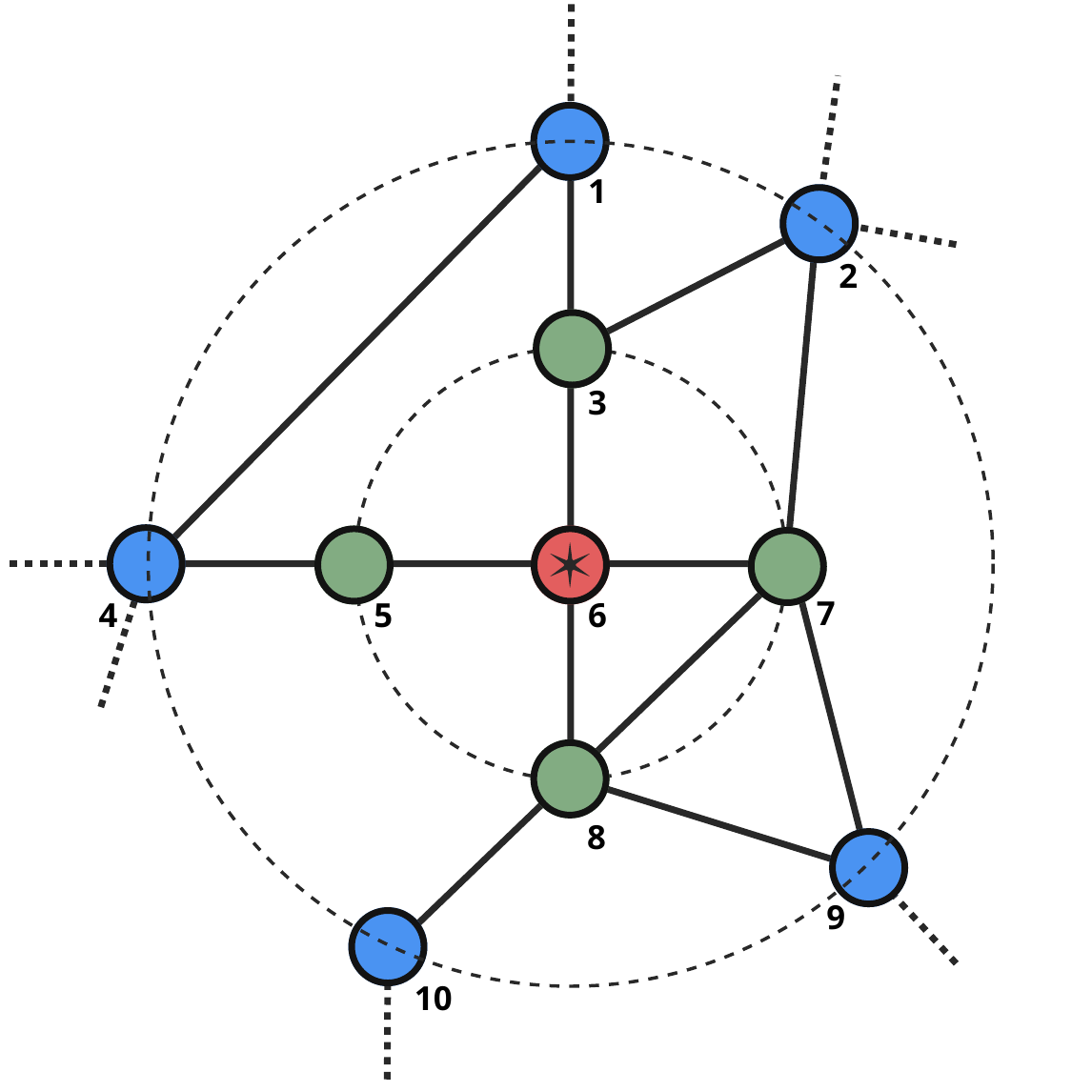}
   \caption{Example of hierarchical degree of \emph{order} $h=2$ for node $6$ (red dot). In this example, the hierarchical degree of \emph{order} $2$ corresponds to the number of links connecting the blue and green dots, resulting in $K_6(2) = 7$.}
  \label{fig:Ex_hierachical}
\end{figure}

Figure~\autoref{fig:Ex_hierachical} shows an example of the calculation of the hierarchical degree for \emph{order} 2. As it is defined, the hierarchical degree will correspond to the number of links between the green dots (layer $R_{1}^6$) and the red dots (layer $R_{2}^6$).

For \emph{order} 1, the hierarchical degree will coincide with the normal degree, so the hierarchical degree can be understood as a natural generalization of the classical node degree. By taking into account larger scale properties of the network, this measurement is able to provide additional information about network hierarchy and connectivity, 
therefore complementing the topological characterization of each node.  Indeed, the topological scale of the respective characterization can be controlled by the choice of the \emph{order} $h$:  the larger the adopted value, the larger the scale of the topological characterization.

\begin{figure}[!htpb]
  \centering
     \includegraphics[width=0.45\textwidth]{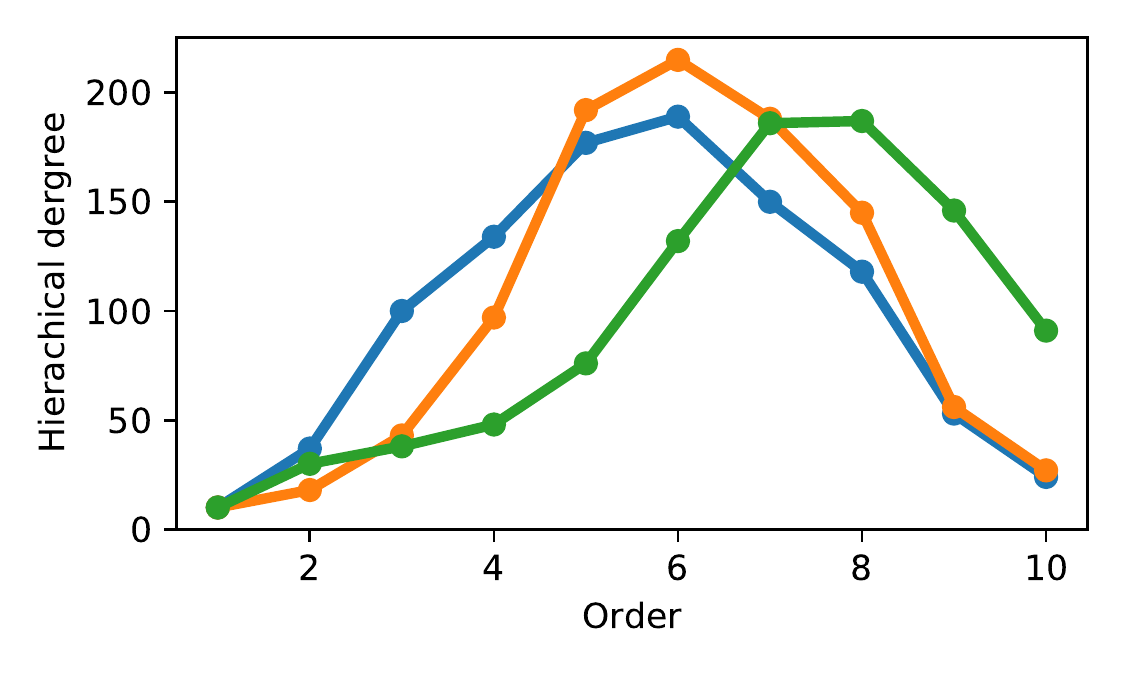}
   \caption{Hierarchical degree, for $h = 1, 2, \ldots 10$, characterizing three arbitrary network nodes having the same traditional degree. These three nodes were obtained from a same BA network with $1,000$ nodes.}
  \label{fig:HD}
\end{figure}

Figure~\autoref{fig:HD} shows the hierarchical degree, for $h = 1, 2, \ldots 10$, of three nodes of a BA network with $1,000$ nodes. 
Though all these three nodes have the same traditional degree equal to 10, their respectively set of hierarchical degrees is substantially different. This example corroborates the fact that there are topological features around each node that are not reflected in the traditional degree. In other words, though the three nodes in this example have substantially different contextual topological properties, they would be understood as having the same features in case only the traditional degree had been taken into account.

\section{Random walks}
A random walk is a kind of stochastic dynamics taking place in a complex
network, consisting of trajectories composed by a succession of aleatory steps taken by a moving agent or walker. Given that several physical world phenomena -- including but not limited to energy and matter diffusion and flow -- are intrinsically related to random walks, several fields of knowledge adopt this tool, including economics (e.g.~\cite{lo2011non}), population genetics(e.g.~\cite{lande1976natural}), physics (e.g.~\cite{risken1996fokker}), and computer science (e.g.~\cite{bar2008random}), to name just a few examples.

\begin{figure}[!htpb]
  \centering
     \includegraphics[width=0.38\textwidth]{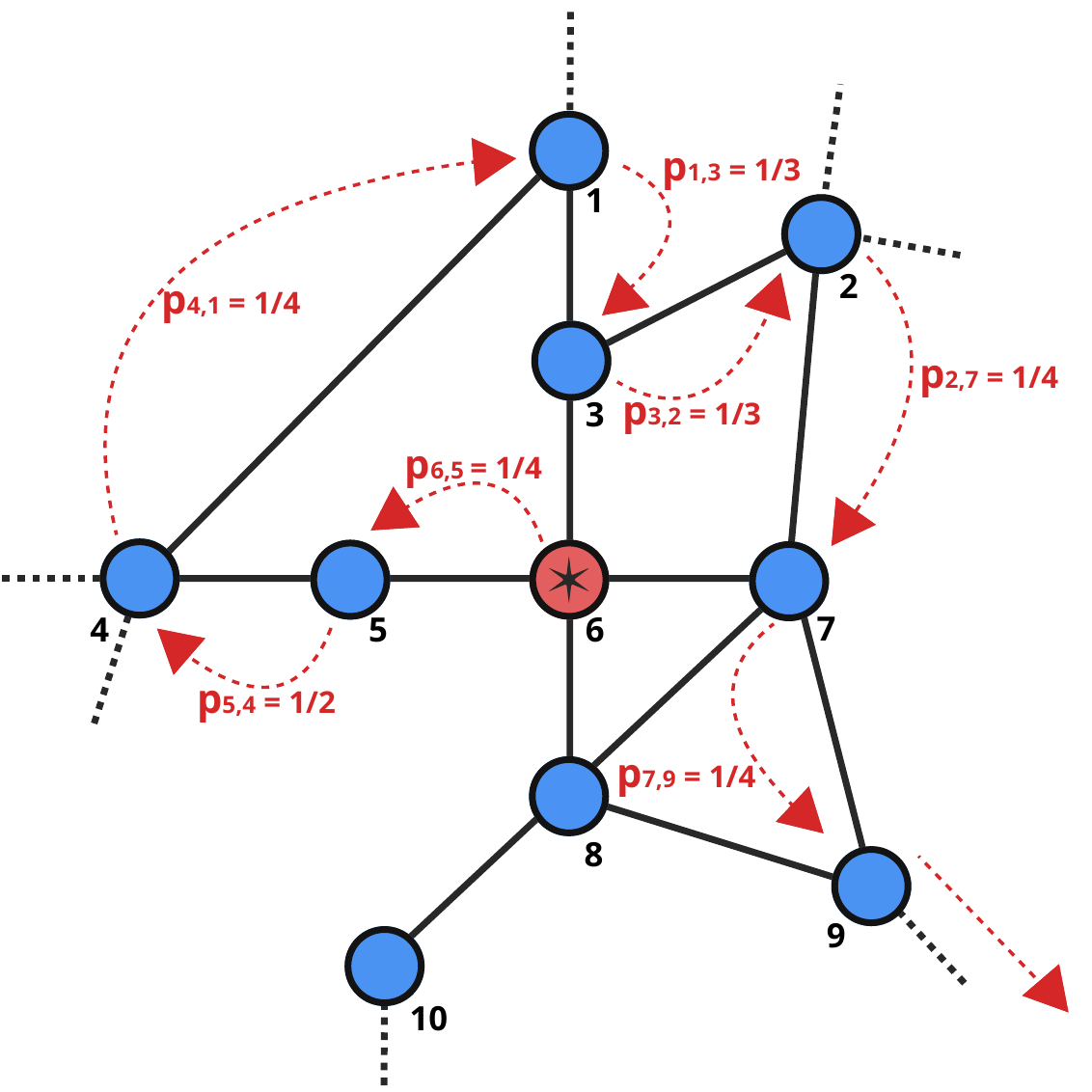}
   \caption{An example of a random walk by an agent on a simple network. The transition probabilities $p_{i,j}$ are also indicated in the figure.}
  \label{fig:random_walk}
\end{figure}

Figure~\ref{fig:random_walk} illustrates a random walk by a single agent on a simple network.  It starts at node $6$ and then, at each successive step, one of the outgoing edges of the current node is chosen with equal probability. This action is repeated until some stopping criteria is verified. The red arrows indicate each of the moves taken along the random walk, therefore defining a contiguous trajectory.  Given that the transition probabilities are all equal at each basic step, this type of random walk is often said to be \emph{uniform}.

As with most dynamic systems, the dynamics of random walks encompasses two successive regimes: (i) \emph{transient}; and (ii) \emph{steady-state} or \emph{equilibrium}. Informally speaking, the latter can be characterized by the system having departed substantially from the respective initial condition and became stable with respect to some attractor, which can be a fixed point or of some other more elaborated type.

There is a virtually infinite number of possible types of random walks. In addition to the discussed above, it is also possible to have the transition probabilities being non-uniform, e.g.~by being proportional to some property of the input/output nodes (e.g.~degree). In addition, as in Levy flights~\cite{shlesinger1995levy}, it is possible to have non-contiguous trajectories, in the sense that the agent can `jump' directly to some more distant node. In addition, random walks can be classified as being deterministic, in which case the movements are performed while optimizing some given property (e.g.~minimal distance).

\section{Entropy}
Informally speaking, in statistical mechanics, entropy is a measurement that quantifies the molecular degree of freedom of a system. Physical entropy is also often associated with randomness, diffusion of matter and energy, and disorder. In information theory, entropy is understood as the amount of information necessary to specify the complete micro-state of the system. 

The Shannon entropy~\cite{balian2004entropy} is a concept originally applied to study the information content of a transmitted message, with possible applications in data compression and communications.

Information entropy is defined as:
\begin{equation}
    H = -\sum_{i=1}^n{\rho_i \log_2(\rho_i)},
\end{equation}
where $p_i$ are a discrete set of probabilities.

All in all, entropy reflects the disorder or heterogeneitu in a system and reflects the uncertainty about these system elements. The $H$ value will be minimum when constant probability distribution, and will be maximum for uniform distributions.

It is also possible to modify the entropy into other related measurements. Of particular interest is the the diversity of a probability density ($D$)~\cite{leinster2012measuring}, defined as the exponential of entropy
\begin{equation}
    D = \exp{(H)}.
\end{equation}

This measurement expresses the effective number of states (values) in the probability density, having applications as in quantifying the diversity of species in ecology~\cite{leinster2012measuring}.

\section{Accessibility}
When applied to characterize the connectivity of networks (graphs) along several neighborhoods~\cite{travenccolo2008accessibility}, the above mentioned diversity has been called accessibility, providing a manner of quantifying the effective number of nodes accessible from a given node in a network (it is also possible to consider the opposite, i.e.~how well the neighbors of a node can access that node). 

As such, the accessibility has some important properties, including: (i) as the hierarchical degree, it takes into account not only the local connectivity around a node, but also the topology along additional neighborhoods (increasing topological scales); (ii) the accessibility considers not only the topology, but also some type of dynamics of interest taking place in the network, which define respective transition probabilities; (iii) it has been found to provide a good indication of the borders of complex networks ~\cite{travenccolo2009border}; (iv) it can be understood as a generalization of the degree and hierarchical node degree incorporating more information about the network topology surrounding a node.

For a given source node $i$, and walks with length $h$ departing from this node, the accessibility can be defined as:
\begin{equation}
    A_i(h) = \exp \left( - \sum_j p_j^{(h)} \log p_j^{(h)} \right),
\end{equation}
where $\{ p_1^{(h)}, p_2^{(h)}, \dots p_j^{(h)}, \dots, p_{N_i(h)}^{(h)} \}$ are the transition probability for reaching the $N_i(h)$ neighbors of $i$ that are at distance $h$.  Observe that transition probabilities can be understood as corresponding to a specific type of \emph{weights} associated to the edges of a network.

\begin{figure}[!htpb]
  \centering
     \includegraphics[width=0.49\textwidth]{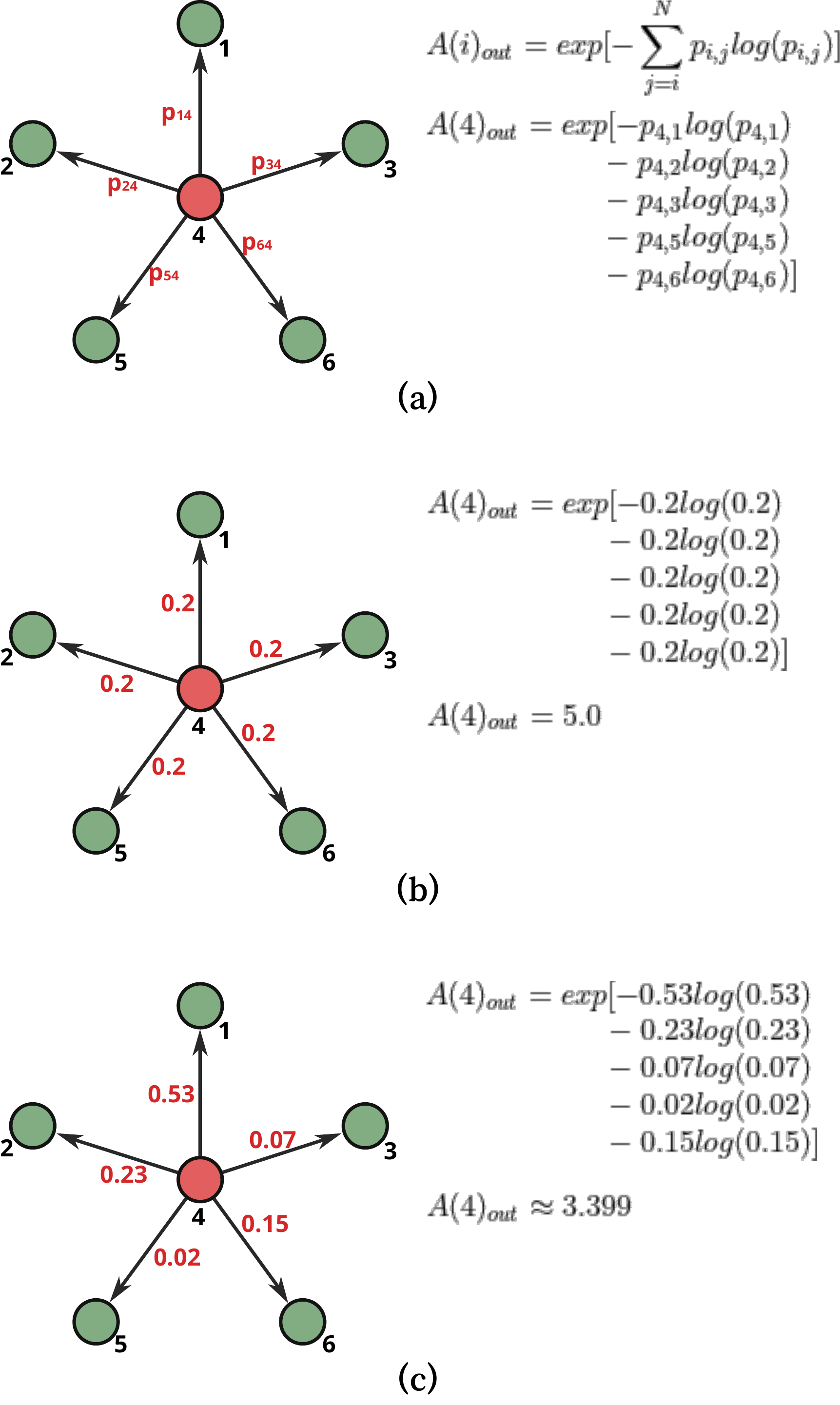}
   \caption{A simple example of how to calculate the accessibility and the impact of the link weights on the result. See text for an explanation.}
  \label{fig:example_out}
\end{figure}

Figure~\ref{fig:example_out} illustrates an example of how to calculate the accessibly of a node in a simple graph, taking into account only the first neighborhood. Figure~\ref{fig:example_out} (a) presents the mathematical definition to be used to calculate the accessibility for a given node of a network; and (b) and (c) depict different configurations of transition probabilities from the reference node to its neighbors.

Having a perfectly uniform distribution of transition probabilities, the situation shown in (b) leads to the maximum accessibility value, which is not verified for the example in (c).  Observe that in the latter situation the nodes will, in the average, be visited in a highly heterogeneous manner during a random walk, with node 1 receiving more than half of the transitions from the reference node 4 (red dot), while
node 5 would be rarely visited.

In summary, the accessibility of a node $i$ given a neighborhood $h$ can be understood to quantify how uniformly the nodes at that neighborhood are visited/accessed from random walks emanating from the reference node.

The accessibility also motivated another directly related topological measurement of complex networks, namely the node symmetry, which involves two aspects: backbone and merged~\cite{silva2016concentric}. As in the accessibility, This measurement aims at considering the interaction of each node with successive neighborhoods, but in a normalized way so as to minimize the effect of the degree of the node of interest.

A software resource has been made available to provide a manner of obtaining the accessibility and symmetry of nodes in a given network. This is provided through the ``network-symmetry'' package available on  Python Package Index (PyPI) repository. The ``network-symmetry'' is a fast library written in C for python to calculate network Accessibility and Symmetry.  The information about these measurements, installing commands, documentation, and steps for its application, is available: \url{https://github.com/ABenatti/network_symmetry} and \url{https://pypi.org/project/network-symmetry/}.

\section{Comparison between the measurements}
In order to compare the accessibility with other measurements, so as to infer how it may be related to and to infer how it may complement them,  here we describe the following experiment.  We created networks according to the following 4 network models with $2,000$ nodes each, all having the same average degree 4.0: 
\begin{enumerate}
    \item Erdős–Rényi model (ER)~\cite{erdos1959random} -- A network obtained from uniformly random connections with probability $p$;
    \item Watts-Strogatz model (WS)~\cite{watts1998collective} -- A random network generation model that provides graphs with small-world properties;
    \item Barabasi-Albert model (BA)~\cite{barabasi1999BA} -- A network characterized by power-law degree distribution;
    \item Geometric model (GEO)~\cite{imai1986efficient} -- Vertices are randomly inserted as points in the 2D unit square, being pairwise connected if they are closer one another by less than the given radius.
\end{enumerate}

\begin{figure}[!htpb]
  \centering
     \includegraphics[width=0.45\textwidth]{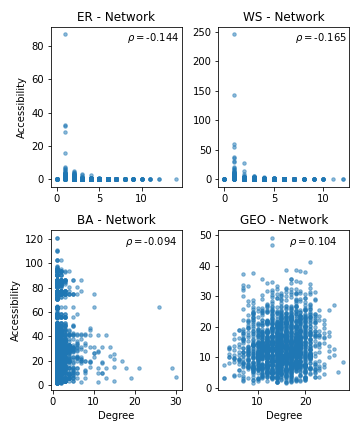}
   \caption{Comparison the degree and accessibility in four network models.}
  \label{fig:degree}
\end{figure}

\begin{figure}[!htpb]
  \centering
     \includegraphics[width=0.45\textwidth]{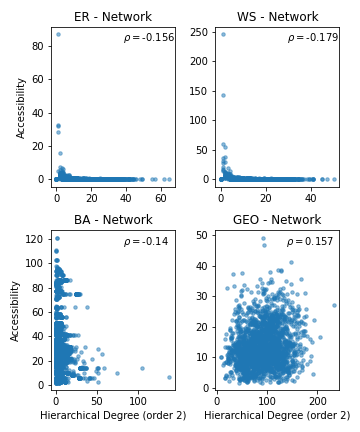}
   \caption{Comparison the hierarchical degree (order 2) and accessibility in four network model.}
  \label{fig:hierarchical}
\end{figure}

\begin{figure}[!htpb]
  \centering
     \includegraphics[width=0.45\textwidth]{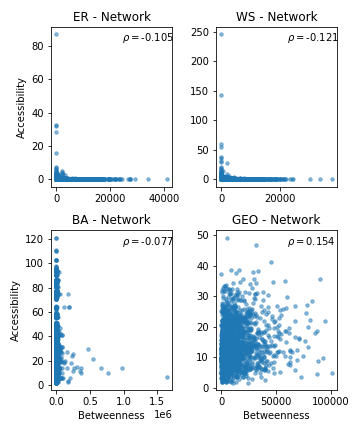}
   \caption{Comparison the betweenness centrality and accessibility in four network model.}
  \label{fig:betweenness}
\end{figure}

\begin{figure}[!htpb]
  \centering
     \includegraphics[width=0.45\textwidth]{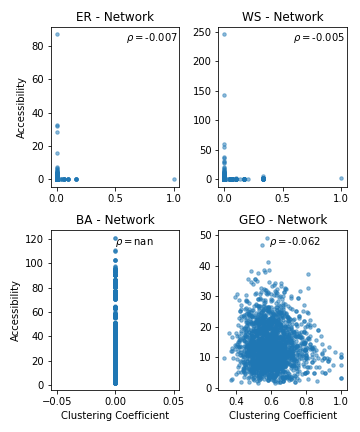}
   \caption{Comparison the clustering coefficient and accessibility in four network model.}
  \label{fig:clustering}
\end{figure}

In order to infer the possible relationship between the accessibility and more traditional topological measurements we obtained the respective scatter plots accompanied by the respective Pearson correlation coefficients ($\rho$), which are shown in Figures~\ref{fig:degree}, \ref{fig:hierarchical}, \ref{fig:betweenness}, and \ref{fig:clustering}

These figures allow a identification of possible relationships between the accessibility with other frequently employed topological measurements, namely with degree, hierarchical degree, betweenness centrality, and clustering coefficient respectively.

In none of the obtained scatterplots can be observed a significant relationship between accessibility and measures of comparison. Therefore, we observe that accessibility is largely unrelated to other more traditional topological properties, therefore providing promising subsidies for complementing the characterization of the structure of complex networks.

\section{Application to border detection}
Accessibility has several applications in network science. One particularly interesting and useful possibility is to use this measurement to propose a formal and objective definition of the borders of a network, therefore allowing the identification of more external or internal nodes~\cite{travenccolo2009border}.  This property reflects the ability of the accessibility also to quantify the centrality of nodes or groups of nodes.

Given a reference node $i$ and a sufficiently large network, the effectiveness of a node $i$ in being accessed from other nodes in a network after $h$ steps depends  on the centrality of the node $i$. In general, nodes near the network border tend to have smaller accessibility values than more central nodes. 

An aspect of special importance regards the fact that the definition of the borders of a  complex network in terms of the accessibility of its nodes also allows a topological scale of particular interest to be selected. For instance, if somebody is particularly interested in the properties of a node given its most immediate neighborhood, a relatively small value of $h$ can be selected.  On the other hand, if the properties of a node need to take into account a more general topological perspective, a higher value of $h$ can be set. It is also possible to perform analyses considering several successive values of $h$, which leads to a multi-scale approach.

\begin{figure*}[!htpb]
  \centering
 \begin{subfigure}[b]{0.4\textwidth}
    \includegraphics[width=\textwidth]{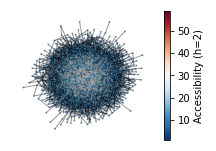}
   \centering (a) ER network
\end{subfigure}
\hspace{0.3cm}
 \begin{subfigure}[b]{0.4\textwidth}
    \includegraphics[width=\textwidth]{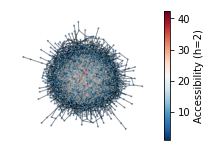}
   \centering (b) WS network
\end{subfigure}
 \begin{subfigure}[b]{0.4\textwidth}
   \includegraphics[width=\textwidth]{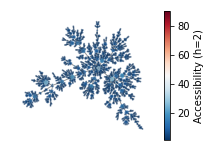}
   \centering (c) BA network
\end{subfigure}
\hspace{0.3cm}
 \begin{subfigure}[b]{0.4\textwidth}
    \includegraphics[width=\textwidth]{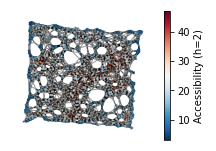}
   \centering (d) GEO network
\end{subfigure}
 \begin{subfigure}[b]{0.4\textwidth}
    \includegraphics[width=\textwidth]{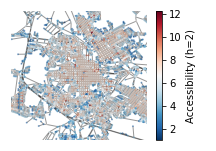}
   \centering (e) streets of São Carlos
\end{subfigure}

   \caption{Accessibility of each node, representing by the colors, of some network models and one real network example.}
  \label{fig:border}
\end{figure*}

Figure~\autoref{fig:border} shows the accessibility of the nodes in some networks examples respective to some well-known models of networks --- ER (a) , WS (b) , BA (c), GEO (d) --- showing the  accessibility values in terms of a heat map for \emph{order} $h=3$.  A real-world network (e), corresponding to the urban streets of São Carlos, SP, Brazil, is also considered respectively to \emph{order} $h=2$.

There are several interesting applications of this method, including the determination of border of urban regions~\cite{travenccolo2009border, tokuda2021spatial}.

\section{A Brief Review of Some Accessibility Applications}

Accessibility as a network measurement has been applied to study urban networks such as in ~\cite{travenccolo2008accessibility} and~\cite{spadon2017identifying}. In the latter work, the authors propose to use topological and geometric measures, characterizing low-access regions, to help identify urban inconsistencies that may occur due to unsuccessfully implemented decisions.

The concept of accessibility has also been explored in order to quantify the influence of individuals concerning the diffusion of information using network dynamics~\cite{pei2013spreading}. This study noted that the neighbors of a node tend to have ability to significantly affect the propagation. 

In order to develop means to identify scientific frauds, the work \cite{amancio2015comparing} considered the hypothesis that artificially generated manuscripts can be distinguished from real scientific papers through topological characterization of complex networks. In this context, accessibility turned out to be useful in specific cases, allowing good performance as a network topological measure to discriminate between real papers from those produced by an automatic generator.

In another work, the accessibility was employed to investigate the possible correlations between the heterogeneous spread of epidemic disease and several attributes of the originating sources~\cite{da2012predicting}. Applying the susceptible–infected–recovered (SIR) model, the authors considered real and theoretical and networks, observing a high correlation between the overall prevalence of the epidemic and the degree, strength, and accessibility of the epidemic sources.

Another application applied the accessibility to quantify the effect of underground systems in facilitating more uniform access to diverse places in Paris e London cities~\cite{costa2010efficiency}. Among the several obtained results, it was shown that the incorporation of public transport tends to enhance the accessibility in cities exhibiting a less uniform topology.

A metric based on accessibility was explored on \cite{das2019accessibility}, considering a network with scientific papers  connected through edges whenever they share the same keyword. This work investigated how well the paper relates with other papers presented in the same conference. The analyses provided a picture of the new emerging topics for future conferences.

The work reported in~\cite{lee2021spatial} proposed to modify the accessibility method to be more specific to road networks. The authors substituted the topological distance with the geometric distance and called this new measurement \emph{access diversity}. The authors concluded that this measurement can identify particularly significant locations in cities that can provide a more intuitive picture of the urban streets.

Network are complex structures that have been employed to represent and investigate a large variety of complex phenomena, from the spread of information to the dissemination of infectious diseases, traffic flow, and the growth of cities, among many other possibilities. The accessibility measurement has been found to provide an interesting resource  for studying these structure.

\section{Conclusions}
\label{sec:conclusion}

The area of network science (e.g.~\cite{barabasi2016network}) has undergone an impressive development since the initial studies of the WWW (e.g.~\cite{montgomery2000trends}) and the Internet (e.g.~\cite{barabasi2001physics}). One topic that has received particular interest regards the characterization of the topological properties of the nodes in these networks. Among these possible measurements, the degree possesses a special relevance as it has been found to be related to several structural and dynamical properties. At the same time, it has been observed (\cite{da2018complex}, see also Fig.~\ref{fig:HD}) that the specification of the degree of each node in a network is not enough to provide a complete representation, in the sense of being invertible. More measurements are required for that finality.

The accessibility has been proposed as a means to generalize the concept of node degree not only along with successive neighborhoods around the node of reference, but also to incorporate specific dynamics of interest. It has been found useful in a large number of applications, some of which revised here, including the definition and identification of the borders of graphs and complex networks.

The present work aimed at providing a hopefully accessible introduction to this measurement, starting by highlighting the special importance of the node degree, and then discussing the related hierarchical degree measurement.  The interesting subject of random walks, which is  frequently adopted while calculating the accessibility, was revised next, followed by a brief presentation of the entropy concept.  With basis on these concepts, the accessibility measurement was then presented and illustrated through simple examples. It was then shown, by considering the correlations between the accessibility and other typically adopted measurements with respect to four network models, that the accessibility presents little relationship with those measurements, therefore being able to provide complementary information.  The possibility to define and obtain the borders of a complex network in a multi-scale manner by using the accessibility was then reviewed and illustrated with respect to four model and one real-world networks.  To conclude our work, we provided a brief review of some of the applications of the accessibility to several different areas.

\section*{Acknowledgements}
Alexandre Benatti thanks Coordenação de Aperfeiçoamento de Pessoal de N\'ivel Superior - Brasil (CAPES) - Finance Code 001. Luciano da F. Costa thanks CNPq (grant no. 307085/2018-0) and FAPESP (grant 15/22308-2).


\bibliography{ref}
\bibliographystyle{ieeetr} 
\end{document}